\documentclass[notitlepage,prd,superscriptaddress,nofootinbib]{revtex4}
 \usepackage{color}
\usepackage{amsmath}
\usepackage{amssymb}
\usepackage{amscd}
\usepackage{latexsym}
\usepackage{hyperref}
\def\a{\alpha}
\def\b{\beta}
\def\g{\gamma}

\renewcommand{\and}{{\quad{\rm and}\quad}}

\def\DH{\rm I\kern-1.5pt\rm H\kern-1.5pt\rm I}

\newcommand{\ba}{\begin{array}}
\newcommand{\ea}{\end{array}}
\newcommand{\be}{\begin{equation}}
\newcommand{\ee}{\end{equation}}
\newcommand{\bea}{\begin{eqnarray}}
\newcommand{\eea}{\end{eqnarray}}
\newcommand{\bi}{\begin{itemize}}
\newcommand{\ei}{\end{itemize}}
\begin{document}
\title{Noncompact $\mathbf{CP}^N$ as a phase space of superintegrable systems}
\author{Erik Khastyan}
\email{khastyanerik@gmail.com}
\affiliation{Yerevan Physics Institute, 2 Alikhanian Brothers St., Yerevan  0036 Armenia}
\author{Armen Nersessian}
\email{arnerses@yerphi.am}
\affiliation{Yerevan Physics Institute, 2 Alikhanian Brothers St., Yerevan  0036 Armenia}
\affiliation{Institute of Radiophysics and Electronics, Alikhanian 1, Ashtarak, 0203, Armenia}
\affiliation{Bogoliubov Laboratory of Theoretical Physics, Joint Institute for Nuclear Research, Dubna, Russia}
\author{Hovhannes Shmavonyan}
\email{hovhannes.shmavonyan@yerphi.am}
\affiliation{Yerevan Physics Institute, 2 Alikhanian Brothers St., Yerevan  0036 Armenia}
\begin{abstract}
We propose the description of superintegrable models with dynamical $so(1.2)$ symmetry, and of the  generic  superintegrable deformations of oscillator and Coulomb systems  in terms of higher-dimensional Klein model (the non-compact analog of complex projective space) playing the role of phase space. We present the expressions of the constants of motion of these systems via Killing potentials  defining the $su(N.1)$ isometries of the K\"ahler structure.
\end{abstract}

\maketitle
\section{Introduction}\noindent
The symplectic manifolds  are the even-dimensional manifolds equipped with closed non-degenerate two-form, which yields the
 non-degenerate Poisson brackets (see, e.g. \cite{arnold}).
In accordance with Darboux theorem any symplectic structure can locally be presented in the canonical form corresponding to the canonical Poisson brackets.
Furthermore, any cotangent bundle  of Riemann manifold can be equipped with the globally defined canonical symplectic structure.
Hence, for the Hamiltonian description of  systems of particles moving on the Riemann space we can restrict ourselves by the canonical symplectic structure (and canonical Poisson brackets).
The non-canonical Poisson brackets are usually used for the description of more sophisticated systems, say, various modifications of tops, (iso)spin dynamics, etc.
An important class of symplectic manifolds are K\"ahler manifolds, which are
 the Hermitian manifolds  whose
 imaginary  parts define the symplectic structures.
K\"ahler manifolds are highly common objects  in almost all areas of theoretical physics, especially in the supergravity and string theory.
However, they usually appear  as configuration spaces of the particles and fields.
Only in  a limited number of physical problems they appear as  phase spaces, mostly for the description of   various versions of  Hall effect, including its higher-dimensional generalizatons (see, e.g. \cite{nair,dolan} and refs therein and to them).
Respectively, the number of the known  nontrivial (super)integrable systems with K\"ahler phase space is very restricted,
 and their  study  is on  the margin of the theory of integrable systems.
 This is especially surprising, given that  quantization of the systems with K\"ahler phase space has been  in the focus of modern geometry  since the invention of the concept of geometric quantization.
 An exceptional integrable model  with K\"ahler phase space  extensively studied  nowadays  is ( compactified ) Ruijesnaars-Schneider model \cite{RS}, but even this system is mostly studied in  canonical coordinates.
 On the other hand, relation of the (existing) integrable systems and of its constants of motion with the isometries of K\"ahler  manifold considered as a phase space
can be useful to understand the geometry of the system, and could be an important  step for   the quantization in    non-canonical coordinates.

 A very simple example of  such system is one-dimensional conformal mechanics formulated  in terms of the Klein model of Lobachevsky plane (``noncompact complex projective plane")  played  the role of phase space \cite{lobach}. Such description, besides elegance,  allows to immediately construct its $\mathcal{N}=2k$ superconformal  extension associated with $su(1.1|k)$ superalgebra.
Similar formulation  of  the higher-dimensional systems  was given in \cite{shmavon0,shmavon}  with the aim to  geometrize (and generalize to higher dimensions)
 the so-called "holomorphic factorization approach"  \cite{ranada} to  the (two-dimensional)  superintegrable oscillator- and Coulomb-like systems invented in \cite{TTW} and extended  to the  spheres and hyperboloids in \cite{acan}.
It was based on the separation of "radial" part of the system (spanned   by  the generators of $so(1.2)$ dynamical symmetry algebra), from the "angular" part, given by  the  Casimir of $so(1.2)$ \cite{angular}.
As a result, the   integrable generalizations of conformal mechanics, and of the (Euclidian, spherical and  hyperbolical) oscillator and Coulomb systems  were formulated   on the {\sl non-K\"ahler} phase space
 $\widetilde{\mathbf{CP}}^1\times \mathcal{M}$, with  $\widetilde{\mathbf{CP}}^1$ being  a Lobachevsky plane ("noncompact complex projective plane") parameterized by  the radial coordinate and momentum, and with $\mathcal{M}$ being the phase space of "angular" part of conformal mechanics. So, the higher-dimensional extension of the  approach  suggested  in \cite{lobach} led to the loss  of $su(1|1)$-symmetric K\"ahler structure of the phase space.
 Nevertheless, a few interesting observations were made there, particularly  the unified  formulation, in terms of complex phase space coordinates, of the hidden symmetry generators of  the maximally superintegrable deformations oscillator- and Coulomb-like systems and of their generalizations to the spheres and hyperboloids \cite{rapid}.

It this paper we  consider   the higher-dimensional  systems  with $su(1.N)$-symmetric K\"ahler phase space which can be considered as a  non-compact version of complex projective space.
The K\"ahler structure  is parameterized by the complex variable $w,\; {\rm Im}\; w \;<0$, and by the $N-1$ complex variables $z^\alpha$, $|z^\alpha|\in [0,\infty), {\rm arg}\; z^\alpha \in [0;2\pi)$. It can be considered as the   $N$-dimensional extension  of the Klein model.
 We connect the complex coordinate $w$ with the radial coordinate and momentum of the conformal-mechanical system spanned by $su(1.1)$ subalgebra, and $z^\alpha$ with  the angular part of that conformal mechanics. Relating  the angular coordinates and momenta with the action-angle variables of the angular part of the integrable conformal mechanics,
  we describe  all symmetries of the generic superintegrable conformal-mechanical systems in terms of the powers of
  the  $su(1.N)$ isometry generators. Then we consider the maximally superintegrable generalizations of the Euclidian  oscillator/Coulomb systems  and express all the symmetries of these superintegrable systems via  $su(1.N)$ isometry generators as well.
Hence, for the systems on Euclidian spaces we perform complete  ``K\"ahlerization" of the construction suggested in \cite{shmavon}.
    Moreover, it seems that in these terms we can construct their  $\mathcal{N}$-extended superconformal extensions, as it was done in \cite{lobach} for one-dimensional case.\\

The paper is organized as follows.

In {\sl Section 2} we present the basic facts on K\"ahler geometry which will be used in our consideration.
 In  {\sl Section 3} we describe the $N$-dimensional generalization of the Klein model
which possesses the $su(1.N)$-invariant K\"ahler structure. Then we constuct, by the use of its isometry generators,
the superintegrable models of conformal mechanics and of  the oscillator- and Coulomb-like  systems.
In {\sl Section 4} we transit to the  canonical coordinates and relate them with the radial coordinate and momentum  spanned by $su(1.1)$ algebra, and with the action-angle variable of its angular part.
We also construct the whole symmetry algebra of the generic  maximally superintegrable deformations  of the oscillator and Coulomb problems, in terms of isometry generators.

\section{Preliminary}\noindent
The symplectic manifold $(M,\omega)$ is  the even-dimensional manifold equipped with closed non-degenerate two-form
\be
\omega=\frac12 \omega_{ij}(x)dx^i\wedge dx^j\;: d\omega=0,\quad \det\omega_{ij}\neq 0.
\ee
This  two-form defines   the non-degenerate Poisson brackets
\be
\{f, g\}=\omega^{ij}(x)\frac{\partial f}{\partial x^i}\frac{\partial g}{\partial x^j}\;,\quad{\rm with}\quad \omega^{ij}\omega_{jk}=\delta^i_k.
\ee
K\"ahler manifold is the  manifold with Hermitian metrics  $ ds^2=g_{a\bar b} dz^a d\bar z^b$ whose
 imaginary  part defines  the symplectic structure
 \be
 \omega_M=\imath g_{a\bar b}dz^a\wedge d\bar z^b,\; d\omega_M=0\quad \Rightarrow\quad g_{a \bar b}dz^ad{\bar z}^b =
\frac{\partial^2 \mathcal{K}}{\partial z^a \partial{\bar z}^{b}}
 \,  dz^ad{\bar z}^b,
 \ee
  where $\mathcal{K}(z,\bar z)$ is a real function (K\"ahler potential)
defined up to holomorphic and antiholomorphic functions: $\mathcal{K}(z,\bar z)\to \mathcal{K}(z,\bar z)+U(z)+{\bar U}(\bar z)$.

Hence, K\"ahler manifold  can be equipped with the  Poisson brackets
\be
\{f,g\}_M=\imath g^{\bar a b}\Big(\frac{\partial f}{\partial {\bar z}^a} \frac{\partial g}{\partial {z}^b}-\frac{\partial g}{\partial {\bar z}^a} \frac{\partial f}{\partial{z}^b}\Big), \quad g^{\bar a b}g_{b \bar c}=\delta^{\bar a}_{\bar c}\; .
\ee
Therefore, the isometries of  K\"ahler structure should preserve both complex and symplectic structures, i.e. they are defined
by  the holomorphic  Hamiltonian vector fields,
 \begin{equation}
{\bf V}_{\mu}=\{h_\mu, \}_M =
    V_\mu^{a}(z)\frac
{\partial}{\partial z^a}+
{\bar V}_\mu^{\bar a}(\bar z) \frac{\partial}
{\partial \bar z^a}, \quad  V^a_\mu=\imath g^{\bar{b}a}\partial_{\bar{b}}h_\mu(z,\bar z) \;.
\end{equation}
The real function $h_\mu (z,\bar z)$   (sometimes  called Killing potential)
obeys the equation
\begin{equation}
\frac{\partial^2 h_\mu}{\partial z^a \partial z^b} -
\Gamma^c_{ab}\frac{\partial h_\mu}{\partial z^c}=0
,
\label{v}\end{equation}
 with $\Gamma^c_{ab} = g^{c\bar{d}}\partial_a g_{b \bar{d}}\,$ being  the  non-vanishing components of the Christoffel symbols.

 The most known examples of nontrivial K\"ahler manifolds are  the $N$-dimensional complex projective space $\mathbf{CP}^N$
 and its  non-compact analog which we will further denote as $\widetilde{\mathbf{CP}}^N$.
 They can be equipped with the $su(N+1)$-invariant (for the compact case) and the $su(N.1)$ invariant (for the non-compact case) K\"ahler metrics, known as the Fubini-Study ones.
 These metrics and respective K\"ahler potentials   are defined by the expressions (with the upper sign corresponding to $\mathbf{CP}^N$, and the lower sign  to $\widetilde{\mathbf{CP}}^N$)
 \footnote{Through this section we use the notation  $z\bar z\equiv \sum_{c=1}^Nz^c \bar z^c$, $zd\bar z\equiv \sum_{c=1}^N z^c d\bar z^c$  etc.}
\be
g_{a\bar b}dz^ad\bar z^b=\frac{gdzd\bar z}{1\pm z\bar z}\mp\frac{g(\bar z dz)(z d\bar z)}{(1\pm z\bar z)^2}, \quad \mathcal{K}=\pm g\log(1\pm z\bar z)
\label{cpn}\ee
where we introduced the positive constant parameter $g>0$ for further convenience.
Let us notice that the  complex projective space $\mathbf{CP}^N$ is defined by the $N+1$ charts, while   its noncompact analog $\widetilde{\mathbf{CP}}^N$ by the single chart.
Moreover, in the latter case the range of validity  of the coordinates $z^a$ is as follows
\be
|z^a| < 1,\quad \sum_{a=1}^N z^a{\bar z}^a < 1
\ee
The inverse metrics defining Poisson brackets is  given  by the expression
$ g^{\bar a b}=\frac1g (1\pm z\bar z)(\delta^{\bar a b} \pm \bar z^a z^b)$,
while the  isometries   are defined by the Killing potentials
\be
h_{a\bar b}=g\frac{\bar z^a z^b \mp \delta_{ a\bar b}}{1\pm z\bar z},\qquad
 h_a=g\frac{2 \bar z^a}{1\pm z\bar z},\qquad {h}_{\bar a}=g\frac{2z^a}{1\pm z\bar z}\;. 
\ee
These generators form the $su(N+1)$ algebra for the upper sign, and the  $su(N.1)$ for the lower one(the generators $h_{a\bar b}$  form $u(N)$ algebra):
\be
\{h_a, h_b\}=0,\quad \{h_a, h_{\bar b}\}=-4\imath h_{a \bar b},\quad
\{h_a, h_{b\bar c}\}=\pm \imath \left( \delta_{ a \bar c}h_b + \delta_{ b\bar c}h_a\right), \quad
\{h_{a\bar b}, h_{c\bar d}\}=\pm  \imath(\delta_{ a\bar d}h_{c \bar b} -\delta_{ \bar b c }h_{a\bar d})
.\label{sun}\ee
\subsection*{Poincar\'e and Klein models of the Lobachevsky plane }
The one-dimensional noncompact complex projective space $\widetilde{\mathbf{CP}}^1$ is  the Lobachevsky plane (upper sheet of two-sheet hyperboloid) proper. Its Fubini-Study metrics results
  in the  $su(1.1)=so(1.2)$-invariant K\"ahler metric parameterized by the  unit disc of two-dimensional plane, which is known as Poincar\'e model \cite{dnf}
\be
ds^2=\frac{g dzd\bar z}{(1-z\bar z)^2},\quad\Rightarrow\quad \mathcal{K}=-g\log(1-z\bar z),\qquad |z|<1.
\ee
In this particular case the Killing potentials read
\be
h= g\frac{1+z\bar z}{1-z\bar z},\quad h_+=g\frac{2\bar z}{1-z\bar z}\;,\quad h_- =g\frac{2z}{1-z\bar z}\; :\quad \{h_+,h_-\}=-4\imath h,\quad\{h_{\pm}, h\}=\mp 2\imath h_{\pm}.
\ee
Performing the transformation
\be
z=\frac{1-\imath w }{1+\imath w }
\ee
we arrive at the so-called Klein model parameterized by upper two-dimensional half-plane \cite{dnf}
\be
ds^2=\frac{gdw d\bar w}{[\imath(w-\bar w)]^2}, \quad \mathcal{K}=-g\log[\imath(w- \bar w)],\qquad {\rm Im}\; w<0.
\ee
The Poisson brackets corresponding to this structure are defined by the relation
\be
\{w, \bar w\}=-\frac{\imath}{g}(w-\bar w)^2,
\ee
while the Killing potentials read
\be
h=g\frac{w\bar w +1 }{\imath(w -\bar w)},\quad h_+=g\frac{(1+\imath w)(1+\imath \bar w)}{\imath(w -\bar w)},\quad h_- =g\frac{(1-\imath w)(1-\imath \bar w)}{\imath(w -\bar w)}\;.
\ee
Instead of these generators it is more convenient to use  their linear combinations
\bea
&H_0=g\frac{w\bar w}{\imath(w-\bar w)}, \quad K_0=\frac{g}{\imath(w- \bar w)},\quad D_0 =g\frac{\bar w +w}{\imath(w-\bar w)}:&\\
&\{K_0,H_0 \}=D_0,\quad \{D_0,H_0 \}=2H_0,\quad \{K_0, D_0\}=2K_0.&
\eea
Introducing the canonical phase space variables $(p,x)$ \cite{lobach}
\be
w=\frac{p}{x}-\imath\frac{g}{x^2}\;:\quad \{x,p\}=1,
\ee
we can represent the Killing potentials in the standard form of the generators of one-dimenisional conformal mechanics \cite{fubini}
\be
 H_0=\frac{p^2}{2}+\frac{g^2}{2x^2}, \quad K_0=\frac{x^2}{2},\quad D_0 =px.
\label{1d}\ee
In the next Section we will extend this mapping to the higher-dimensional noncompact complex projective space.

\section{Noncompact complex projective space: Klein model}
Let us construct  $N$-dimensional  analog of the Klein model from the
Fubini-Study structure of noncompact complex projective space
$\widetilde{\mathbf{CP}}^N$  given by the expressions \eqref{cpn} with a lower sign.
For this purpose we perform the transformation
\be
z^N=\frac{1-\imath w}{1+\imath w}, \quad z^\alpha=\sqrt{2}\frac{\tilde z^\alpha}{1+\imath w},
\label{dlp}\ee
which yields the following expressions for the K\"ahler structure and  potential (here and further instead of $\tilde z^\alpha$ we use the former notation $z^\alpha$)
\be
ds^2=
\frac{g[dw+\imath\bar z^\a  dz^\a ][d\bar w - \imath z^\b d \bar z^\b )]}{[\imath(w-\bar w)- z^\gamma \bar{{z}}^\gamma]^2}+\frac{g dz^\a d\bar z^\a }{\imath(w-\bar w)- z^\gamma \bar{{z}}^\gamma} ,
\ee
\be \mathcal{K}=-g\log\left[{\imath(w-\bar w)- z^\gamma \bar{{z}}^\gamma}\right],\qquad \alpha,\beta,\gamma =1,\ldots N-1,
\label{NKlein}\ee
with the following  range of validity  of  the coordinates $w, z^\alpha$
\be
{\rm Im}\; w<0,\qquad \sum_{\alpha=1}^{N-1}z^\alpha {\bar z}^\alpha< -2\; {\rm Im}\; w.
\label{rd}\ee
The  respective Poisson brackets  are defined by the relations
\be
\{w,\bar w\}=-{A}(w-\bar w),\quad \{w, \bar z^\alpha\}={A}\bar z^{\a},\quad \{z_\alpha,\bar z_\beta\}=\imath{A}\delta^{\bar \b \a},
\ee
where
\be
A:=\frac{\imath(w -\bar w)- z^\gamma\bar z^\gamma}{g}\;.
\ee
The Killing potentials of the  K\"ahler structure \eqref{NKlein} are defined by the expressions
\be
h_{N \bar N}=\frac{w\bar w +1}{ A},\quad h_{\a \bar N}=\frac{1}{\sqrt{2}}\frac{ \bar z^\a (1-\imath w) }{ A },
\quad h_{\a\bar \b}=\frac{\bar z^\a z^\b + \frac{1}{2}\delta_{ \a \bar\b}(1+\imath w)(1-\imath \bar w)}{ A}
\label{KpK0}\ee
\be
h_{N}=\frac{(1+\imath w)(1+\imath \bar w)}{A },\quad h_\a=\sqrt{2}\frac{\bar z^\a(1+\imath w)}{ A}\quad
.
\label{KpK}\ee
These potentials form $su(N.1)$ algebra, which in the given notation  reads the same as in \eqref{sun} with a lower sign and $a=N,\a$.
Below  we will refer  to this representation as
the $N$-dimensional  Klein model.

For our purposes, instead of   Killing  potentials \eqref{KpK0},\eqref{KpK} it is more convenient to use the following ones
\be
H=\frac{w\bar w}{A},\quad K=\frac{1}{ A},\quad D=\frac{w+\bar w}{ A},\quad
H_{\alpha \bar N}=\frac{\bar z^\alpha w}{A}, \quad H_{\alpha}=\frac{\bar z^\alpha}{ A},\quad H_{\alpha\bar\beta}=
\frac{\bar z^\alpha z^\beta}{A}.
\ee
Certainly, these functions  are not independent, for  there are many obvious relations between them, e.g.
\be
H=\sum_{\alpha=1}^{N-1}\frac{ H_{\alpha \bar N} {\bar H}_{N \bar\alpha}}{H_{\alpha\bar\alpha}},\quad H_{\alpha\bar\beta}=\frac{H_\alpha H_{\bar\beta}}{K},\quad {\rm etc}.
\ee
In these terms the
  $su(1.N)$ algebra relations read
\bea
&\{ H , K \}=-D, \quad \{H , D \}=-2H, \quad \{K , D \}=2K,&\label{0}\\
&\{H , H_\a \}=-H_{\a \bar N}, \quad \{H , H_{\a \bar N}\}=\{H , H_{\a \bar \b}\}=0,
&\label{44}\\
&
\{K , H_{\a \bar N}\}=H_\a, \quad\{K , H_\a \}=\{K , H_{\a \bar \b}\}=0,
&\label{45}\\
&
\{D, H_\a \}=-H_\a, \quad \{D , H_{\a \bar N}\}=H_{\a \bar N}, \quad \{D , H_{\a \bar \b} \}= 0,
&\\
&\{H_\a ,H_\b \}= \{H_{\a \bar N},H_{\b \bar N}\}=\{H_{\alpha},H_{\beta \bar N}\}=0,&\\
&\{H_\a , H_{\bar \b}\}= -\imath K\delta_{\a \bar \b},\quad
\{H_{\a \bar N}, H_{N \bar \b} \} = -\imath H \delta_{ \a \bar\b}, \quad
\{H_{\a \bar \b}, H_{\g \bar \delta}\}=\imath(H_{\a \bar \delta} \delta_{\g \bar\b}-H_{\g \bar \b}\delta_{ \a \bar\delta}),&\\
&
\{H_\a , H_{N \bar \b}\}= H_{\a \bar \b} +\frac{1}{2}\left(g+\sum_\gamma H_{\gamma \bar\gamma}-\imath D\right)\delta_{ \a \bar\b},
&\\
&
\{H_\a , H_{\b \bar \g} \}=-\imath H_\b \delta_{\a \bar \g}, \quad
\{H_{\a \bar N},H_{\b \bar \g} \} = -\imath H_{\b \bar N} \delta_{ \a \bar\g}.&\label{infty}
\eea
So, the generators $H,K,D$ define the conformal algebra $su(1.1)=so(1.2)$, and the generators $H_{\alpha\bar\beta}$  define  the algebra $u(N-1)$.\\

It is seen that
\begin{itemize}
\item
the Hamiltonian  $H$  has   two sets of constants of motion $H_{N\alpha}$ and $H_{\alpha\bar\beta}$ (see \eqref{44}), therefore it defines  superintegrable system;

\item
the Hamiltonian  $K$  has   two sets of constants of motion as well,  $H_{\alpha}$ and $H_{\alpha\bar\beta}$ (see\eqref{45}). Thus, it defines the superintegrable system as well;

\item the triples $(H, H_{N\alpha}, H_{\alpha\bar\beta})$ and  $(K, H_{\alpha}, H_{\alpha\bar\beta})$ transform  into each other  within
 discrete transformation
\be
(w, z^\alpha)\to (-\frac1w,\frac{z^\alpha}{w})\quad\Rightarrow D\to -D, \quad\left\{\begin{array}{c}(H, H_{N\alpha}, H_{\alpha\bar\beta})\to (K, -H_{\alpha}, H_{\alpha\beta}),\\
(K, H_{\alpha}, H_{\alpha\beta})\to (H,- H_{N\alpha}, H_{\alpha\bar\beta})
\end{array}\right. .
\label{duality}\ee
\end{itemize}
Adding  to the  Hamiltonian $H$  the appropriate function of $K$, we get the superintegrable oscillator- and Coulomb-like systems.

\subsubsection*{Oscillator-like Hamiltonian}\noindent
We define the oscillator-like  Hamiltonian by the expression (cf.\eqref{1d})
\be
H_{osc}=H+\omega^2K
\label{osc}\ee
and introduce the following generators
\be
 A_{\alpha}= H_{\alpha\bar N}+ \imath\omega H_{\alpha},\quad
 B_{\alpha}= H_{\alpha\bar N}- \imath\omega H_{\alpha} \; :
\qquad \left\{
\begin{array}{c}\{H_{osc},A_\alpha\}=- \imath \omega A_{\alpha}\\
\quad \{H_{osc},B_\alpha\}= \imath \omega B_{\alpha}
\end{array}
\right.
.
\label{AB}\ee
These generators and their complex conjugates form the following
algebra
\bea
&\{A_{\alpha},\bar A_{\beta}\}=-\imath \big(H_{osc}-\omega( g+\sum_{\gamma=1}^{N-1}  H_{\gamma \bar \gamma} )\big)\delta_{\alpha \bar \beta}+2 \imath \omega H_{\alpha \bar \beta},
&\\
&\{B_{\alpha},\bar B_{\beta}\}=-\imath \big(H_{osc}+\omega( g+\sum_{\gamma=1}^{N-1}  H_{\gamma \bar \gamma} )\big)\delta_{\alpha \bar \beta}-2 \imath \omega H_{\alpha \bar \beta},
&\\
&\{A_{\alpha},\bar B_{\beta}\}=-\imath \delta_{\alpha \bar \beta}\big( H_{osc}-2\omega^2 K + \imath\omega D \big),
&\eea
with their Poisson brackets with $H_{\alpha \bar\beta}$ reading
\bea
&\{A_{\alpha},H_{\beta\bar\gamma}\}=-\imath \delta_{\alpha \bar \gamma}A_{\beta},\qquad \{B_{\alpha},H_{\beta\bar\gamma}\}=-\imath \delta_{\alpha \bar \gamma } B_{\beta}\qquad
.&
\eea
Then we immediately deduce that the Hamiltonian \eqref{osc} besides $H_{\alpha\bar\beta}$, has the  additional constants of motion which provide the system by the maximal superintegrability property
\be
M_{\alpha\beta}=A_{\alpha} B_{\beta}= H_{\alpha\bar N}H_{\beta\bar N}+\omega^2 H_\alpha H_\beta+\imath \omega(H_\alpha H_{\beta\bar N}-H_{\alpha\bar N}H_{\beta})=\frac{\bar{z}^\alpha \bar{z}^{\beta}}{A^2}(w^2+\omega^2) \;:\quad \{H_{osc}, M_{\alpha\beta}\}=0.\label{M}\ee
For sure, these constants of motion are functionally dependent, so that among them one can choose the  $N-1$ integrals which
 guarantee superintegrability of the system, e.g.  $P_\alpha\equiv P_{\alpha\alpha}$ only,  like  in \cite{shmavon}.
The  generators \eqref{M} and the $su(N)$ generators $H_{\alpha\beta}$  form the following symmetry algebra
\be
\{H_{\alpha\bar\beta}, M_{\gamma\delta}\}=\imath \delta_{\bar \beta \gamma}M_{\alpha \delta}+\imath \delta_{\bar \beta\delta}M_{\gamma \alpha},
\qquad
\{M_{\alpha\beta}, M_{\gamma\delta}\}=0,
\ee
\subsubsection*{ Coulomb-like Hamiltonian}\noindent
We define the  Coulomb-like   Hamiltonian  with the   additional constants of motion which provide  the system  by the maximal superinetgrability property as follows (cf. \eqref{1d})
\be
H_{Coul}=H-\frac{\gamma}{\sqrt{2 K}},  \quad  R_{\alpha}=H_{\alpha\bar  N}+\imath \gamma\frac{ H_{\alpha }}{(g+\sum_{\gamma=1}^{N-1} H_{\gamma \bar \gamma})\sqrt{2 K}}\;:\quad\{H_{Coul},R_{\alpha}\}=\{H_{Coul},H_{\alpha\bar \beta}\}=0.
\label{Coul}\ee
The whole  symmetry algebra   is as follows
\be
\{R_\alpha,R_{\bar\beta}\}=-\imath\delta_{\alpha \bar \beta}\Bigg(H_{Coul}-\frac{\imath\gamma^2}{2(g+\sum_{\gamma=1}^{N-1} H_{\gamma \bar \gamma})^2}\Bigg)+\frac{\imath\gamma^2 H_{\alpha\bar\beta}}{2(g+\sum_{\gamma=1}^{N-1} H_{\gamma \bar \gamma})^3},\quad\{H_{\alpha\bar\beta},R_\gamma\}=\imath \delta_{\gamma \bar \beta}R_{\alpha},\quad
\{R_\alpha,R_\beta\}=0.
\ee
To clarify the origin of these models it is convenient to transit  to the canonical coordinates.

\section{Canonical coordinates}\noindent
For the introduction of the canonical coordinates
 we transit  from the complex coordinates to the real ones
\be
 w=x+\imath y , \quad z^\alpha = q_\alpha {\rm e}^{\imath\varphi_\alpha}, \quad{\rm where}\qquad y <0,\quad  q_\alpha\geq 0,\quad \varphi_\alpha\in [0, 2\pi ),\quad q^2:=\sum_{\alpha=1}^{N-1} q_\alpha^2 < -2y.
\ee
Then
we write down
  the symplectic/K\"ahler one-form   and identify it with the canonical one
\be
\mathcal{A}=-\frac{g}{2}\frac{dw+d{\bar w} -\imath(z^\alpha d{\bar z}^\alpha - {\bar z}^\alpha d z^\alpha )}{{\imath(w-\bar w)- z^\gamma \bar z^\gamma}}:=p_xdx  + \pi_\alpha d\varphi_\alpha.
\ee
This yields the following expressions for the canonical coordinates and momenta,
\be
p_x=g\frac{1}{2y+q^2},\quad \pi_{\alpha}=-g\frac{q^2_\alpha}{2y+q^2}\quad\Leftrightarrow\quad q_\alpha=\sqrt{-\frac{\pi_\alpha}{p_x}},\quad y=\frac{\pi+g}{2p_x},\quad{\rm with}\quad \pi:=\sum_{\alpha=1}^{N-1}{\pi_\alpha}
\ee
Thus, the complex coordinates are expressed via canonical ones as follows
\be
w= x+\imath\frac{\pi+g}{2p_x},\qquad z^\alpha=\sqrt{-\frac{\pi_\alpha}{p_x}}{\rm e}^{\imath\varphi_\alpha}.
\ee
For the complete analogy with
one-dimensional case \cite{lobach} we perform further canonical transformation
$(x,p_x)\to (p_r/r,-r^2/2)$  and re-write
the above expression in a more convenient form
 \be
w= \frac{p_r}{r}-\imath\frac{\pi+g}{r^2}, \qquad z^\alpha=\frac{\sqrt{2\pi_\alpha}}{r}{\rm e}^{\imath\varphi_\alpha},\qquad{\rm with}\qquad  r>0,\quad \pi_\alpha \geq 0,\quad \varphi_\alpha\; \in[0, 2\pi).
\ee
and
\be
A=\frac{\imath(w -\bar w)- z^\g\bar z^\g}{g}=\frac{2}{r^2}.
\ee
In these terms the generators of conformal algebra \eqref{0} take the form of conformal mechanics with separated "radial" and "angular" parts (cf. \cite{angular}),
\be
H=\frac{p^2_r}{2} +\frac{\mathcal{I}}{r^2},\quad K= \frac{r^2}{2},\quad D= p_r r,
\label{HKD}\ee
 where the angular part of Hamiltonian  is  given  by the expression
\be
\mathcal{I}=\frac12\left(\sum_{\alpha=1}^{N-1}\pi_\alpha + g\right)^2
\;.
\label{angular}\ee
The rest generators of $su(1.N)$ algebra read
\be
H_{\alpha \bar N}=\sqrt{2\pi_\a}\left(\frac{p_r}{2}-\imath\frac{\pi+g}{2r}\right){\rm e}^{-\imath \varphi_\a},
\quad H_\alpha=r\sqrt{\frac{\pi_\a}{2}}{\rm e}^{-\imath\varphi_\a},
\quad H_{\alpha \bar\beta}=\sqrt{\pi_\a \pi_\b}{\rm e}^{-\imath\left(\varphi_\a-\varphi_\b\right)},
\ee
with the basic Poisson brackets $\{r,p\}=1$ and $\{\varphi_\a,\pi_\a\}=1$.

In these  coordinates the oscillator- and Coulomb-like Hamiltonians \eqref{osc},\eqref{Coul} take the  form,
\be
H_{osc}=\frac{p^2_r}{2} +\frac{\mathcal{I}}{r^2}+\frac{\omega^2r^2}{2},\quad H_{Coul}=\frac{p^2_r}{2} +\frac{\mathcal{I}}{r^2}-\frac{\gamma}{r},
\label{oscCoul}\ee
with $\mathcal{I}$ given by \eqref{angular}.

The generic conformal mechanics   with the angular part $\mathcal{I}_{gen}(\pi,\varphi)$ can be defined via $su(1.N)$ generators  by the expression
\be
H_{gen}=H+\frac{\mathcal{I}_{gen}(H_\alpha/\sqrt{K},H_{\bar\beta}/\sqrt{K})-(\sum_{\gamma=1}^{N-1} H_{\gamma\bar\gamma}+g)^2}{2K}.
\ee
However, we are mostly interested in the study of integrable and superintegrable systems.
Thus, we have to restrict ourselves by the  particular cases of angular Hamiltonians.

 \subsection*{Superintegrable Systems}\noindent
In accordance with Liouville theorem, the integrability of the system with $2N$-dimensional phase space means the existence $N$ functionally independent involutive integrals $F_1=H,\ldots, F_N :\{F_a,F_b\}=0$. This yields the existence of the so-called action-angle variables $(I_{a}(F), \Phi_a)$:
\be
H=H(I), \quad \{I_a,\Phi_b\}=\delta_{ab},\quad \{I_a,I_b\}=\{\Phi_a,\Phi_b\}=0,\qquad \Phi_a\in [0, 2\pi),\qquad  a,b=1,\ldots, N.
\ee
The system becomes maximally superintegrable when the Hamiltonian is expressed via action variables as follows
\be
H=H\left(\sum_{a=1}^N n_aI_a\right),\quad n_a\in \mathcal{N}
\ee
where $n_a$ are integers (or rational numbers). Indeed, in that case the system possesses the
additional (non-involutive) integrals $I_{ab}=\cos({n_a\Phi_b-n_b\Phi_a})$, among them $N-1$ integrals are functionally independent.

Now, let us suppose that
 $\pi_\alpha,\varphi_\alpha$ are related with the  action-angle variables $(I_\alpha,\Phi_\alpha)$ of some $(N-1)$-dimensional angular mechanics by the relations
\be
\pi_\alpha=n_\alpha I_\alpha,\quad \varphi_\alpha=\frac{\Phi_\alpha}{n_\alpha},\qquad {\rm where}\quad n_\alpha\in \mathcal{N}.
\label{paa}\ee
Upon this identification the angular Hamiltonian \eqref{angular} takes a form
 \be
 \mathcal{I}=\frac12\left(\sum_{\alpha=1}^{N-1}n_\alpha I_\alpha +g\right)^2,\qquad{\rm with}\quad n_\alpha\;\in\;\mathcal{N},
 \label{angularGen}\ee
This is precisely the class of angular Hamiltonians  which provides
the superintegrable generalizations of the conformal mechanics, and of the oscillator and Coulomb  systems on the $N$-dimensional Euclidian  spaces \cite{rapid}!

Though the relations \eqref{0}-\eqref{infty} hold upon this identification, the generators $H_{\alpha},H_{\alpha\bar N}, H_{\alpha\bar\beta}$ become  locally defined,
 $\varphi_\alpha\;\in \;[0, 2\pi/m_\alpha)$, so  they fail to be constants of motion.
However, taking their  relevant powers  we  get the globally defined generators which form the nonlinear algebra
\bea
&&{\widetilde H}_\alpha :=(H_{\alpha})^{n_\alpha}=d_\a (I) r^{n_\alpha} {\rm e}^{-\imath\Phi_\alpha}, \\
&&{\widetilde H}_{\alpha\bar N} := (H_{\alpha\bar N})^{n_\alpha}= d_{\alpha\bar N}(I)\left(p_r-\imath\frac{\sum_{\gamma=1}^{N-1} n_\gamma I_\gamma+g}{r}\right)^{n_\alpha}{\rm e}^{-\imath \Phi_\a},\\
&&{\widetilde H}_{\alpha\bar\beta} :=(H_{\alpha\bar\beta})^{n_\alpha n_\beta}=d_{\alpha\bar\beta}(I) {\rm e}^{-\imath\left(n_\b\Phi_\a-n_\a\Phi_\b\right)},
\eea
where
\be
d_{\alpha}(I)=\left(\frac{n_\alpha I_\alpha}{2}\right)^{n_\a /2},\quad d_{\alpha\bar N}(I)=\left(\frac{n_\alpha I_\alpha}{2}\right)^{n_\a /2},\quad d_{\alpha\bar\beta}(I)=({n_\a n_\b I_\a I_\b})^{n_\alpha n_\beta /2}.
\label{d}\ee
Thus, we get
\be
\{H,{\widetilde H}_{\alpha\bar N}\}=\{H,{\widetilde H}_{\alpha\beta}\}=0,\qquad \{K,{\widetilde H}_{\alpha}\}=\{K,{\widetilde H}_{\alpha\beta}\}=0,
\ee
where $H, K$ are defined by \eqref{HKD} and \eqref{angularGen}.
For sure, the functions \eqref{d}, being dependent on action variables only, do not affect the commutativity of the additional integrals with the Hamiltonian.\\

In a similar way we   construct  the constant of motion of the oscillator- and Coulomb-like systems given  by \eqref{osc},\eqref{angularGen} and  \eqref{Coul},\eqref{angularGen}, respectively.

For the oscillator-like system \eqref{osc} the integrals take the form
\be
{\widetilde M}_{\alpha\beta}:=(A_{\alpha}B_{\beta})^{n_\alpha n_\beta}=\frac12 d_{\alpha\bar\beta}(I) {\rm e}^{-\imath\left(n_\b\Phi_\a-n_\a\Phi_\b\right)} \left(\left( \imath p_r+\frac{\sum_{\gamma=1}^{N-1}n_\gamma I_\gamma+g}{r}\right)^2-\omega^2r^2 \right)^{n_\a n_\b},
\ee
with $A_\alpha, B_\beta$ given by \eqref{AB}, \eqref{paa}.

For the Coulomb-like system \eqref{Coul} the integrals take the form
\be
\widetilde{R}_\alpha= (R_{\alpha})^{n_\alpha}=d_\a (I) {\rm e}^{-\imath\Phi_\alpha}\left(p_r+\frac{\imath \gamma }{\sum_{\gamma=1}^{N-1}n_\gamma I_\gamma +g}-\frac{\imath \left(\sum_{\gamma=1}^{N-1}n_\gamma I_\gamma +g\right)}{r}\right)^{n_1}
\ee

There are a few interesting simple systems whose angular parts are given by \eqref{angularGen} with $g\neq 0$, among them are,
\begin{itemize}
\item "charge-monopole" system (and respective systems with oscillator/Coulomb potentials),
\be
\mathcal{H}=\sum_{a=1}^3\frac{p^2_a}{2} +\frac{s^2}{2r^2},\qquad \{p_a,x^b\}=\delta_a^b,\quad\{p_a,p_b\}=s\frac{\varepsilon_{abc}x^c}{r^2},\quad\{x^a,x^b\}=0.
\ee
Its angular part is defined by the (two-dimensional)  spherical Landau problem,  with the following Hamiltonian (see, e.g.\cite{sagatel}, where one can find the expressions for action-angle variables for the angular part)
\be
\mathcal{I}=\frac12 \Big(I_1+I_2 +|s|\Big)^2,\qquad I_{1,2}\in[0, \infty),
\ee
with $s$ being the monopole number.

\item Smorodinsky-Winternitz system
\be
H_{SW}=\sum_{a=1}^N\Big(\frac{p^2_a}{2}+\frac{g^2_a}{2x^2_a}+\frac{\omega^2x^2_a}{2}\Big).
\ee
The angular Hamiltonian of this system is given by the expression \eqref{angularGen} with  (see, e.g. \cite{galajinsky})
\be
k_\alpha= 2\omega , \quad g=\sum_{a=1}^{N-1}|g_a|.
\ee
For sure, this system could be viewed as a trivial case of rational Calogero model, which also belongs to the class of systems above.

\item Rational Calogero model associated  with Coxeter  root system \cite{calogero-root}
 ${\mathcal R} \subset {{R}}^N$,
\begin{equation}
\label{qHCox}
{{\cal H}}_{Cal} =  \sum_{a=1}^{N} \frac{ p_a^2}{2}  +
\sum_{\alpha\in {\mathcal R}_+} \frac{g^2_\alpha(\alpha\cdot \alpha)}{2(\alpha \cdot x)^2}\ ,\qquad
\{{ p}_a, x_b\}= \delta_{ab}
\end{equation}
where $g_\alpha\ge 0$ is  a Weyl-invariant multiplicity function on the set of roots \cite{Humphreys}.

The spectrum of the  angular part of quantum rational Calogero model was found in  \cite{flp}.  Taking its classical limit,  one can get the expression of the angular (part of)
  generalized rational Calogero model  in terms of action variables \cite{rapid}. It   given by \eqref{angularGen}, with
$n_\alpha$ being  the degrees of the basic homogeneous Weyl-invariant
polynomials,  and  $g=\sum_{\alpha\in\mathcal R_{+}}g_\alpha $.
\end{itemize}
Let us notice that in the angular Hamiltonian \eqref{angular} the nonzero constant $g\neq 0 $ appears, and the range of validity of the action variables  is fixed to be  $I_\alpha\in [0,\infty)$.
As a result, the standard oscillator and Coulomb systems cannot be included in the proposed description, since  for these systems we should choose $g=0, I_\alpha \in [0,\infty)$.
The first condition leads to the vanishing of K\"ahler structure and Poisson brackets,  while the absorbtion of constant $g$ by the action variables immediately
yields the change of the range of validity of the action variables.
However, a minor complication allows to involve in  our picture  the generic superintegrable conformal mechanics, oscillator and Coulomb systems as well.

Using the expressions of the constants of motion   presented in \cite{shmavon},  we can immediately write down the constants of motions of those  systems written in terms of Killing potentials.
\begin{itemize}
\item {\sl Conformal mechanics}
\be
\mathcal{H}=H-\frac{g(g+2\sum_{\gamma=1}^{N-1} H_{\gamma\bar\gamma})}{4K},\qquad \mathcal{H}_{\alpha\bar N}:={H}_{\alpha\bar N}+\imath\frac{g\bar{z}^\alpha}{2}={H}_{\alpha\bar N}+\imath g\frac{H_\alpha}{2K}\; :\quad\{ \mathcal{H}, \mathcal{H}_{\alpha\bar N}\}=\{ \mathcal{H}, {H}_{\alpha\bar\beta}\}=0.
\ee
\item {\sl Oscillator-like system}

\be
\mathcal{H}_{osc}=H_{osc}- \frac{g(g+2\sum_{\gamma=1}^{N-1} H_{\gamma\bar\gamma})}{4K},\qquad
 \{\mathcal{H}_{osc}, H_{\alpha\bar\beta}\}= \{\mathcal{H}_{osc}, \mathcal{A}_{\alpha} \mathcal{B}_{\beta}\}=0
 \ee
where
\be
\mathcal{A}_\alpha={H}_{\alpha\bar N}+\imath g\frac{H_\alpha}{2K}+\imath \omega H_\alpha, \quad \mathcal{B}_\alpha={H}_{\alpha\bar N}+\imath g\frac{H_\alpha}{2K}-\imath \omega H_\alpha
\ee
\item {\sl  Coulomb-like system}
\be
\mathcal{H}_{Coul}=H_{Coul}- \frac{g(g+2\sum_{\gamma=1}^{N} H_{\gamma\bar\gamma})}{4K},\qquad \{\mathcal{H}_{Coul}, \mathcal{R}_{\alpha} \}=0
\ee
where
\be
 \mathcal{R}_{\alpha}=R_\alpha +\imath g\frac{H_\alpha}{\sqrt{2 K}}\Big( \frac{1}{\sqrt{2K}}+\frac{\gamma}{(g+\sum_{\gamma=1}^{N} H_{\gamma\bar\gamma})\sum_{\gamma=1}^{N} H_{\gamma\bar\gamma}} \Big)
 \ee
\end{itemize}
The transition to the action-angle variables \eqref{paa} is obvious.

Hence, we have shown how to   describe the  superintegrable deformations of oscillator and Coulomb systems in terms of noncompact complex projective spaces $\widetilde{\mathbf{CP}}^N$.

\section{Concluding remarks}
In this paper  we have  shown that the superintegrable generalizations of conformal mechanics, oscillator and Coulomb systems can  be naturally described in terms of the noncompact complex projective space considered as a phase space. This observation yields some  interesting directions for further studies, among them
\begin{itemize}
\item the construction of the $\mathcal{N}=2k$ superconformal mechanics  associated with $su(1.N|k)$ superalgebra.
For this purpose one should consider phase superspace equipped with the K\"ahler structure with the potential
\be
\mathcal{K}=-g\log (\imath( w-\bar w)-z^\a \bar z^\a -\imath\eta_A\bar\eta_A),\quad A=1,\ldots k,
\ee
where $\eta_A$ are Grassmann variables.
This should be the direct generalization of the one-dimensional system considered in \cite{lobach}. We expect that it will be possible to construct, in a similar way, the
$\mathcal{N}=2k$  supersymmetric extensions of the considered oscillator-  and (repulsive) Coulomb-like systems as well, in particular, the superextension of Smorodinsky-Winternitz system.

\item Performing the transformation to the higher-dimensional Poincare model via \eqref{dlp}, we expect to present  the considered models in  the  Ruijsenaars-Schneider-like form
 and in this way to find, some superinegrable cases of the Ruijsenaars-Schneider systems, as well as their supersymmetric/superconformal extensions.

\item  describing the superintegrable deformations of the free particle on the spheres/hyperboloids, and  the  spherical/hyperbolic oscillators, in a similar way. For this purpose
we expect to consider the "$\kappa$-deformation" of the K\"ahler structure of the Klein model, in the spirit of
 the so-called ``$\kappa$-deformation approach" developed in \cite{ranada}.
\item constructing  spin-extensions of the above models, choosing the noncompact analogs of complex Grassmanians as  phase spaces.
\end{itemize}
We plan to address  these problems soon.

\acknowledgements We thank Tigran Hakobyan for his  interest in our work and  useful comments.
This work was supported by the Armenian Science Committee  within joint research project with
Russian Foundation for Basic Research,  grant  20RF-023.
 The work of E.Kh. was completed  within the Regional Doctoral Program on Theoretical and Experimental Particle Physics Program
sponsored by VolkswagenStiftung, and the  ICTP Affiliated Center Program AF-04.

\end{document}